\documentclass[superscriptaddress,secnumarabic,twocolumn,
 amssymb,amsmath,nobibnotes,aps,prd,showkeys,showpacs]{revtex4}
\usepackage{graphicx,epsfig,amssymb,amsmath,amstext}

\usepackage{graphicx}
\usepackage{bm}

\begin{document}

\title{Comment on ``Chaotic orbits for spinning particles in Schwarzschild spacetime''}

\author{Georgios Lukes-Gerakopoulos}
\email{gglukes@gmail.com}
\affiliation{Institute of Theoretical Physics, Faculty of Mathematics and Physics, 
 Charles University in Prague, 18000 Prague, Czech Republic}

\begin{abstract}
 The astrophysical relevance of chaos for a test particle with spin moving in
 Schwarzschild spacetime was the objective of \cite{Verhaaren10}. Even if the
 results of the study seem to be qualitatively in agreement with similar works,
 the study presented in \cite{Verhaaren10} suffers both from theoretical and
 technical issues. These issues are discussed in this comment. 
\end{abstract}

%
%
\maketitle

\section{The issue of scaling} \label{sec:scale}

The Mathisson-Papapetrou (MP) equations \cite{Mathisson37,Papapetrou51} describe
the motion of an extended body in the pole-dipole approximation on a curved
spacetime. In the MP description the body has some internal degrees of freedom,
which are constraint to fix its centroid, i.e. the wordline along which the body
moves. This constraint is imposed when a spin supplementary condition (SSC) is
chosen. In \cite{Verhaaren10} the Tulczyjew (T) SSC has been chosen. This makes
\cite{Verhaaren10} comparable with previous similar works since T SSC has been
used in \cite{Suzuki97} (Schwarzschild background as in \cite{Verhaaren10})
and in \cite{Hartl03a,Hartl03b,Han08} (Kerr background). 

When studying the MP equations with T SSC one chooses the mass of the test
particle to be described by the contraction of the four-momentum, i.e.,
$P^a P_a=-{\cal M}^2$, since this mass is a conserved quantity for T SSC
(see, e.g., \cite{Semerak99}). In \cite{Verhaaren10} the mass is chosen with
respect to the four-velocity, i.e. $P^a V_a=-\mu$, which is not a conserved 
quantity for T SSC (see, e.g., \cite{Semerak99}). Choosing the one mass over
the other is a matter of what ones sets as an observer. Since in T SSC we 
observe with respect to the four-momentum $P^a$ is somehow not ``natural'' to
measure the mass with respect to $V^a$. Apart from being conceptually strange to
choose $\mu$ as the mass for the MP with T SSC, this brings along some further
complications when the spin is scaled with respect to $\mu~m$ as stated to be
done in \cite{Verhaaren10}, where $m$ is the mass of the central black hole
(the notation of \cite{Verhaaren10} is adopted). For example, the measure of the
spin is a constant of motion for T SSC (see, e.g., \cite{Semerak99}), but when
one normalizes the spin with something that varies, the constancy of the spin
ceases to be the case.

In order to understand these complications, let us discuss the scaling issue
in more detail. The spin is usually scaled with respect to $m~{\cal M}$,
or with respect to ${\cal M}^2$ to make it a dimensionless quantity (see, e.g.,
discussion in \cite{Hartl03b}). The MP equations can be written in scale free
units if we use the $m~{\cal M}$ scaling for spin as  
 \begin{eqnarray}   \label{eq:MPeqsNorm} 
  \frac{D~P^\mu/{\cal M}}{d \tau/m} &=-\frac{1}{2}~({R^\mu}_{\nu\kappa\lambda} m^2)
  V^\nu \frac{S^{\kappa\lambda}}{m {\cal M}} ~~,\nonumber \\
  \frac{D~S^{\mu\nu}/(m {\cal M})}{d \tau/m} &=(P^\mu~V^\nu-V^\mu~P^\nu)/{\cal M}~~, 
 \end{eqnarray}
where each quantity was written with respect to its scale factor. It is easy to 
see that the scale factors cancel out. Now, if we follow the scalings 
suggested in \cite{Verhaaren10}, then
 \begin{eqnarray}   \label{eq:MPeqsNormV} 
  \frac{D~P^\mu/{\cal M}}{d \tau/m} &=-\frac{1}{2}~({R^\mu}_{\nu\kappa\lambda} m^2)
  V^\nu \frac{S^{\kappa\lambda}}{m \mu} ~~,\nonumber \\
  \frac{D~S^{\mu\nu}/(m \mu)}{d \tau/m} &=(P^\mu~V^\nu-V^\mu~P^\nu)/{\cal M}~~, 
 \end{eqnarray}
and we get
 \begin{eqnarray}   \label{eq:MPeqsNormVf} 
  \frac{D~P^\mu}{d \tau}=-\frac{{\cal M}}{2\mu}~{R^\mu}_{\nu\kappa\lambda}
  V^\nu S^{\kappa\lambda} ~~,\nonumber \\
  \frac{D~S^{\mu\nu}}{d \tau}=\frac{\mu}{{\cal M}}(P^\mu~V^\nu-V^\mu~P^\nu)~~,
 \end{eqnarray}
where the scales do not vanish. 

One could argue that the scales would vanish if the momentum $P^a$ was scaled
with respect to $\mu$ and not with respect to ${\cal M}$. This is true, but in
\cite{Verhaaren10} it is said that $P^a P_a=-1$, which suggests either that the
momentum in \cite{Verhaaren10} is scaled with respect to ${\cal M}$ or that
$P^a P_a=-{\cal M}^2/\mu^2=-1$. The latter cannot be the case since during the
evolution $\mu$ varies, while ${\cal M}$ is a constant, and in general 
$\mu\neq{\cal M}$ for T SSC (see, e.g., \cite{Semerak99}). The rescaling of the
momentum with respect to ${\cal M}$ is reflected on the eqs.~(10),~(13),~(14)
in \cite{Verhaaren10}. What is missing from eqs.~(13),~(14) is the rescaling
of the spin four-vector $S^a$. $S^a$ is the vector counterpart of $S^{ab}$ 
see, e.g., eq.~(10) in \cite{Verhaaren10}. Eqs.~(13),~(14) hold for the
$m {\cal M}$ rescaling of the spin (e.g., the usual rescaling for T SSC used in
\cite{Hartl03a}). But, if the $m \mu$ rescaling of the spin was used in 
\cite{Verhaaren10}, as stated in Sec.~II of \cite{Verhaaren10}, then the
corresponding eqs.~(13),~(14) in \cite{Verhaaren10} should include the ratios
$\mu/{\cal M}$ as shown in the corresponding eqs.~\eqref{eq:MPeqsNormVf} shown
above. Thus, the rescaling implied by eqs.~(13),~(14) in \cite{Verhaaren10} is
inconsistent with the $m \mu$ rescaling of the spin stated in \cite{Verhaaren10}.


This inconsistency is reflected also on the eqs.~(15--17) of \cite{Verhaaren10}.
Furthermore, in eq.~(15) of \cite{Verhaaren10} the first term $P^a$ in the
numerator should not share the denominator with the second term. Namely, eq.~(15)
should read
\begin{align}
 \label{eq:v_p_TULb}
 V^\mu = \frac{\mu}{{\cal M}^2} \left(P^\mu - 
          \frac{~^*R^{*\nu\rho\kappa\lambda} S_\rho P_\kappa S_\lambda}
          {{\cal M}^2+~^*R^{*\alpha\beta\gamma\delta} S_\alpha P_\beta S_\gamma P_\delta }
          \right)  \qquad .
\end{align}
But, this is probably just a typo. The main issue here is that the stated
rescaling is in contradiction with the formulas presented.

Note also that if $\mu$ was considered constant, we would not be able to
normalized the four-velocity so that $V_a~V^a=-1$ in order to evolve the MP
equations with T SSC. The variability of $\mu$ is what allows the four-velocity
normalization (see, e.g., \cite{Semerak99}).

\section{The Poincar\'{e} section and the Lyapunov number issue}

Poincar\'{e} sections are a useful tool to discern chaos from order in a
two degrees of freedom Hamiltonian system. Such a Hamiltonian system, for instance,
corresponds to geodesic motion in an axisymmetric and stationary spacetime
background. Regular orbits are represented by closed zero width smooth curves,
while chaotic orbits are represented by scattered points covering a non-zero
width space on the section. So, the Poincar\'{e} sections shown in Figs.~(1),~(2)
of \cite{Verhaaren10} should indeed represent regular orbits. 

However, the MP equations with T SSC have not been yet been described by a
canonical Hamiltonian formalism (contrary to what is stated on page 3 in
\cite{Verhaaren10}), and the spin introduces more phase space dimensions to the
system than the geodesic spinless case. So, one should be careful when
interpreting 2D~``Poincar\'{e}'' sections for a test particle with spin moving
on a Schwarzschild background. In the latter system what one gets usually for
regular orbits on a 2D section are projections of tori which dimensionality is
higher than three. These projections on a 2D section are represented by no zero
width curves in the case of regular orbits, and there is no straightforward way
to discern chaos from order just from inspecting these 2D sections. Thus, from
Figs.~(3),~(4) of \cite{Verhaaren10} one cannot tell whether the orbits are
regular or not contrary to what is stated in \cite{Verhaaren10} (see, e.g., 
caption of Fig.~4 in \cite{Verhaaren10}). Actually, in \cite{Han08} there is a
case (Fig.~3 in \cite{Han08}), where an orbit looking like those in
Figs.~(3),~(4) of \cite{Verhaaren10} was characterized as ``chaotic mimic'',
because when such 2D section was tested with other indicators of chaoticity,
the other indicators implied that the orbit was regular.

One such indicator of chaoticity is the characteristic Lyapunov number $\lambda$.
Lyapunov numbers are used to cross check the results of Figs.~(3),~(4) in
\cite{Verhaaren10}. However, defining the Lyapunov number for curved spacetimes
is not a straightforward task as discussed in Sec.~III of \cite{Verhaaren10}.
There is the issue of the time, and of the deviation vector ${\bar \xi}$. The
proper time $\tau$, used in \cite{Verhaaren10}, is the time usually employed for
trajectories in curved spacetime to solve the time issue. But, the deviation
vector is a more complicated problem. This vector is defined in a tangent space
of the phase space. More precisely this tangent space is a collection of tangent
spaces along the points consisting the trajectory. There is even a difficulty to
comprehend what exactly this tangent space means in the general relativistic
setup. But apart from the latter issue, the main problem with the deviation
vector is what is the norm of this vector. Different norms $\xi=|\bar \xi|$
result in different values for the Lyapunov number. It was found, however, that
the sign of the Lyapunov number should not be affected by the choice of the norm.
In \cite{Verhaaren10} it is stated that for simplicity the Euclidean norm was
preferred as the norm for $\xi$ in their work, but no further informations about
the explicit form of the norm are provided. Namely, there are the questions of
how was the Euclidean norm applied in the Schwarzschild coordinates of the orbit,
and how was the spin incorporated in the Euclidean norm of the deviation vector.
Without the above explanations the results of this work are not reproducible and
ambiguous.

A standard way to find whether an orbit is chaotic or not by using Lyapunov 
numbers is the $\ln \lambda$ vs. $\ln \tau$ plot (see, e.g., \cite{Han08}).
For a regular orbit the deviation vector grows linearly, i.e. $\xi \propto \tau$,
which means that $\lambda \propto \frac{\ln \tau}{\tau}$. On the logarithmic
plot this implies that for a regular orbit
$\displaystyle \lim_{\tau \rightarrow 0}\ln \lambda \rightarrow -\infty$, i.e.
$\lambda \rightarrow 0^+$, with a slop equal to $-1$. Note that the Lyapunov
number is practically evaluated for finite time, and during this even for 
regular orbits $\lambda>0$. As long as $1/\lambda$ is of the order of magnitude
of the time $\tau$ we have evolved the orbit, we cannot tell whether an orbit 
is chaotic or not. For a chaotic orbit the deviation vector grows exponentially
$\xi \propto e^{\lambda \tau} $, which means that in the logarithmic plot we get
a constant value $\ln \lambda$ after the Lyapunov time $\tau_\lambda=\lambda^{-1}$
is reached. In order to be sure that one gets a chaotic orbit, one has to evolve
the orbit at least for two orders of magnitude more after $\tau_\lambda$ is
reached do $\lambda$ is not any more comparable with $1/\tau$.

In \cite{Verhaaren10} the above standard procedure is absent. The 
procedure to find the Lyapunov number in \cite{Verhaaren10} is based on a
phenomenological model (eq~(28) in \cite{Verhaaren10}) which is irrelevant with
the basic principles describing the evolution of the deviation vector discussed
in the above paragraph. The example in Fig.~6 of \cite{Verhaaren10} evolves an
orbit for $\tau=10^5$ and predicts an orbit with a Lyapunov number
$\lambda \approx 3.787~10^{-4}$. For
$\tau=10^5$ $\lambda \propto \frac{\ln \tau}{\tau}\approx 6.47~10^{-4}$. 
Namely, for the amount of time the orbit has been evolved in Fig.~6 the Lyapunov
number has a value that is comparable with a value of $\lambda$ corresponding to
a regular orbit. Thus, one cannot tell safely whether the orbit is chaotic or not.  
In fact, there are many Lyapunov--like chaotic indicators (see, e.g., \cite{Skokos10}
for a review), but none indicator can safely reveal the chaotic nature
of an orbit at time scales comparable with the Lyapunov time. All the indicators
show the nature of the orbit, much after the magnitude of the Lyapunov time has
been reached. 

In order to investigate the dependence of the chaoticity of the MP equations on
the spin's value in \cite{Verhaaren10}, the energy, the angular momentum, the initial
radius $r$ of the orbit and the orientation of the spin are kept constant, while
the spin's value varies (Figs.~7--13,~15 in \cite{Verhaaren10}). This might seem
reasonable since the investigation depends only on one varying parameter, but
this approach is misleading. The phase space of the system is mixed in the sense
that chaotic and regular orbits coexist in the phase space. When we change a
parameter of the system, the phase space changes and as a consequence the
position of the orbit we suppose to follow changes as well. For example, if we
start with an initial setup at which the orbit we examine is chaotic, by
changing the spin parameter the orbit with the same otherwise setup will
correspond to another trajectory, which might be chaotic or not. Even if we
assume that the method of estimating the Lyapunov numbers followed in
\cite{Verhaaren10} was correct, then what we see in Figs.~7--13,~15 of
\cite{Verhaaren10} is not correlated with the chaoticity of one single orbit,
i.e. it cannot provide qualitative informations about the development of the
system. If one wanted to do such analysis, the only possible way to follow an
orbit through the changing phase space is to track it down in the frequency
domain. For regular orbits their characteristic frequencies are their
identification numbers. For chaotic orbits the unstable periodic orbits and
their corresponding asymptotic manifolds are defining the domain in the phase
space which a chaotic orbit covers, so one should track the unstable periodic 
orbits. 

\begin{acknowledgments}
 GL-G is supported by UNCE-204020 and GACR-14-10625S
\end{acknowledgments}

\end{document}